\documentclass[epjST]{svjour}

\usepackage[dvips]{graphicx}

\newcommand{\epg}{e p \to e p \gamma}
\newcommand{\psg}{\gamma^{(*)} p \to  \gamma p}
\newcommand{\plltt}{P_{LL}-P_{TT}/\epsilon}
\newcommand{\plt}{P_{LT}}
\newcommand{\pll}{P_{LL}}
\newcommand{\ptt}{P_{TT}}
\newcommand{\ale}{\alpha_E}
\newcommand{\bem}{\beta_M}
\newcommand{\lama}{\Lambda_{\alpha}}
\newcommand{\lamb}{\Lambda_{\beta}}
\newcommand{\nucleonmass}{M}

\begin{document}

\title{
Real and Virtual Compton Scattering: the nucleon polarisabilities
}


%
\author{E.~J.~Downie\inst{1} \and H.~Fonvieille\inst{2} }
\institute{ Institut f\"{u}r Kernphysik, Johannes Gutenberg-Universit\"{a}t Mainz, 55099 Mainz, Germany
\and
Clermont Universit\'{e},  Universit\'{e} Blaise Pascal, CNRS/IN2P3, Laboratoire de Physique Corpusculaire, BP 10448, F-63000 Clermont-Ferrand, France 
}
%

%
%
%
%
\abstract{
We give an overview of low-energy Compton scattering  \ $\psg$ \ with a real or virtual incoming photon. These processes allow the investigation of one of the fundamental properties of the nucleon, i.e. how its internal structure  deforms under an applied static electromagnetic field. Our knowledge of nucleon polarisabilities and their generalization to non-zero four-momentum transfer will be reviewed, including the presently ongoing experiments and future perspectives.
} 

\maketitle
%


%
%

\section{Introduction} \label{sec-intro}

The internal structure of the nucleon has been a matter of intensive research for more than five decades. It can be studied in a clean way using electromagnetic probes, i.e. a real or virtual photon of small enough wavelength. The simplest electromagnetic process, $\gamma^* N \to N$, probes the charge and magnetization densities inside the nucleon via the measurement of elastic form factors  $G_E^N$ and $G_M^N$. Next in complexity comes the Compton scattering reaction, with either a real or virtual incoming photon: \ $\gamma^{(*)} N \to \gamma N$. At low energy the structure-dependent part of this process is parameterised by the polarisabilities, which describe the deformation of the nucleon structure in an external quasi-static electromagnetic field and are therefore fundamental characteristics of the nucleon. These observables are sensitive to the full excitation spectrum of the nucleon, contrary to the elastic form factors.
The aim of this article is to give a synthetic overview of our present knowledge of nucleon polarisabilities, with an emphasis on experimental results and ongoing experiments. The first part will be devoted to Real Compton Scattering (RCS) and the second to Virtual Compton Scattering (VCS). 
%
%
A theoretical review of recent developments in the field can be found in \cite{pasquini-crc}.


\section{Real Compton Scattering} \label{sec-rcs}

\subsection{Scalar polarisabilities and their study with Real Compton Scattering}

When one studies the classical properties of a given material, for
instance Carbon, one sees that many attributes are intrinsic, for example
density or electron shell structure, but many, such as the Young's
Modulus, thermal conductivity or the magnetic ordering describe the
response of the material to an external constraint placed upon it.
The nature of these responses results from the internal structure and
bonding of the material and their study can give us insight into its
constituents and internal configuration. When one looks at the Particle
Data Group's list of proton properties (see table \ref{tab:ProtonProps}),
most of the known quantities: spin, parity, mass etc., are intrinsic
and only two of the listed properties describe such a response to
a constraint. These are the scalar polarisabilities: $\ale$,
the electric polarisability and $\bem$, the magnetic polarisability.
These quantities can be interpreted as the response of the proton
structure to the application of an external electric or, respectively,
magnetic field.

\begin{table}
\begin{center}
\caption{\label{tab:ProtonProps}Proton properties as listed by the Particle
Data Group \cite{PDG:2010}. }

\begin{tabular}{|c|c|}
\hline 
Mass  & $938.272013\pm0.000023\, MeV$\tabularnewline
\hline 
Charge  & +1\tabularnewline
\hline 
$I(J^{p})$  & $\frac{1}{2}(\frac{1}{2}^{+})$\tabularnewline
\hline 
Charge radius  & $0.8768\pm0.0069\, fm$\tabularnewline
\hline 
Mean life  & $>5.8\cdot10^{29}$ years\tabularnewline
\hline 
Magnetic moment  & $2.792847356\pm0.000000023\mu_{N}$\tabularnewline
\hline 
Electric dipole moment  & $<0.54\cdot10^{-23}\, e \cdot cm  $\tabularnewline
\hline 
Valence quarks  & uud\tabularnewline
\hline 
Electric polarisability $\ale$  & $12\pm 0.6 \cdot10^{-4}fm^3$ \tabularnewline
\hline 
Magnetic polarisability $\bem$  & $1.9\pm0.5\cdot10^{-4}fm^3$ \tabularnewline
\hline
\end{tabular}
\end{center}
\end{table}

In order to understand the implications of the currently accepted
scalar polarisability values, $\ale= (12\pm 0.6) \cdot10^{-4}fm^3$
and $\bem=(1.9\pm0.5)\cdot10^{-4}fm^3$
\cite{PDG:2010}, 
it is helpful to be aware of the typical sizes of
such quantities for other physical systems. When one considers that
a perfectly conducting sphere has polarisabilities which are of the
order of one quarter of its volume and a Hydrogen atom one tenth,
it becomes clear that the response of the nucleon to a static electromagnetic
field is extremely small \cite{Drechsel:2002ar}. This tells us that
the proton is a very rigid object due to the strong binding of its
constituents. This is not only true of the proton, but also the neutron
and pion, as can be seen in table~\ref{tab:Hadron_Pol}, both
of which are also strongly bound hadronic objects.

\begin{table}
\begin{center}
\caption{\label{tab:Hadron_Pol}Polarisabilities of various hadrons.}

\begin{tabular}{|c|c|c|c|}
\hline 
Hadron  & Source  & $\bar{\alpha}\,(10^{-4}fm^{3})$  & $\bar{\beta}\,(10^{-4}fm^{3})$\tabularnewline
\hline 
Proton  & \cite{OlmosdeLeon:2001zn} $(\bar{\alpha}+\bar{\beta})$ fixed  & $12.1\pm0.4\mp1.0$  & $1.6\pm0.4\mp0.8$\tabularnewline
\hline 
Proton  & \cite{OlmosdeLeon:2001zn} $(\bar{\alpha}+\bar{\beta})$ free  & $11.9\pm0.5\mp1.3$  & $1.2\pm0.7\mp0.3$\tabularnewline
\hline 
Proton  & \cite{Schumacher:2005an} & $12.0\pm0.6$  & $1.9\mp0.6$\tabularnewline
\hline 
Neutron  & \cite{Schumacher:2005an} & $12.5\pm1.7$  & $2.7\mp1.8$\tabularnewline
\hline 
Pion ($\pi^{+}$)  & \cite{Ahrens:2004mg} & \multicolumn{2}{c|}{ $( \bar{\alpha} - \bar{\beta} ) =11.6\pm3.4$ }\tabularnewline
\hline
\end{tabular}
\end{center}
\end{table}

In order to access the polarisabilities, it is necessary to have an
electromagnetic field in which one can place the object of study.
For the nucleon, this can be found in the case of 
Real Compton Scattering (RCS),
for which the unpolarised differential cross section at low energies
can be expressed as follows: 
%
%
\begin{eqnarray}
\begin{array}{lll}
\left(\frac{d\sigma}{d\Omega}\right)=\left(\frac{d\sigma}{d\Omega}\right)_{Point}-\\
\omega\omega'\left(\frac{\omega'}{\omega}\right)^{2}\frac{e^{2}}{m}\left[\frac{\bar{\alpha}+\bar{\beta}}{2}(1+\cos\theta)^{2}+\frac{\bar{\alpha}-\bar{\beta}}{2}(1-\cos\theta)^{2}\right] +O(\omega^{3}) \\
\label{Eqn:RCS}\\
\end{array}\end{eqnarray} 
%
%
where $\omega$ and $\omega'$ are the (lab) initial and final photon energies,
respectively, and $\theta$ is the scattering angle of the photon
in the lab system.  $\left(\frac{d\sigma}{d\Omega}\right)_{Point}$
is the cross section for scattering from a point object, containing
the static properties of the proton, its mass, charge and anomalous
magnetic moment \cite{Wissmann:2004}.  As in the above equation, one often 
sees $\ale$ written as $\bar{\alpha}$ and $\bem$ written as $\bar{\beta}$,
this means only that it was measured in the dynamic process of 
Compton Scattering, rather than with a static electromagnetic field
but despite the differing notion, both quantities are identical \cite{Schumacher:2005an}. Using eq.(\ref{Eqn:RCS}), it
is possible to extract $\ale$ and $\bem$ by analysing
the differential cross section of RCS,
at low incoming photon energy as a function of angle.

\subsection{The TAPS Measurement at MAMI}

\begin{figure}
\centering\includegraphics[width=10cm]{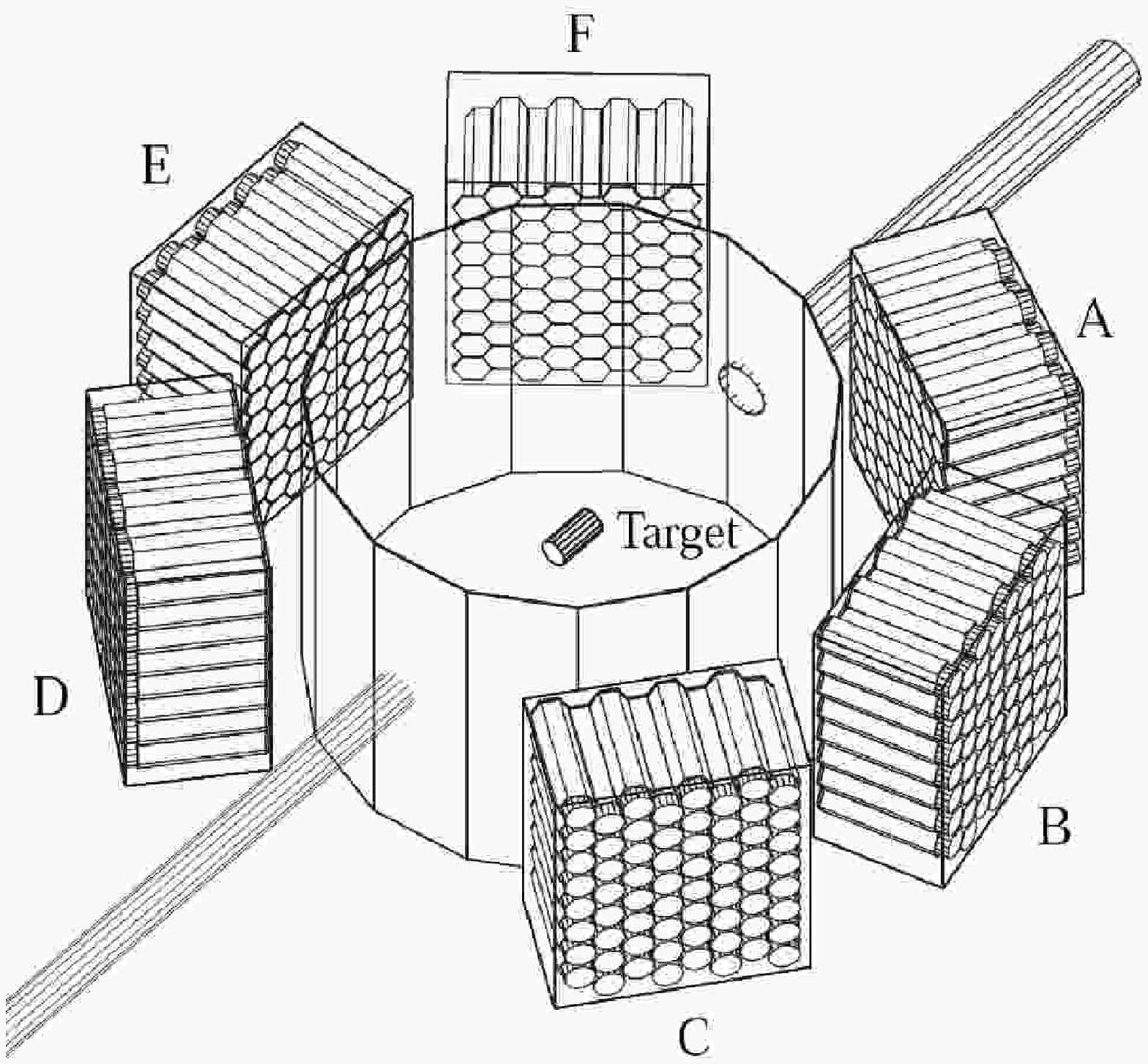}
\caption{\label{fig:TAPS_Setup}Experimental setup of the TAPS measurement
at MAMI. The photon beam enters the diagram at the upper right hand
side \cite{OlmosdeLeon:2001zn}.}
\end{figure}

\begin{figure}
\centering
\includegraphics[width=5.3cm]{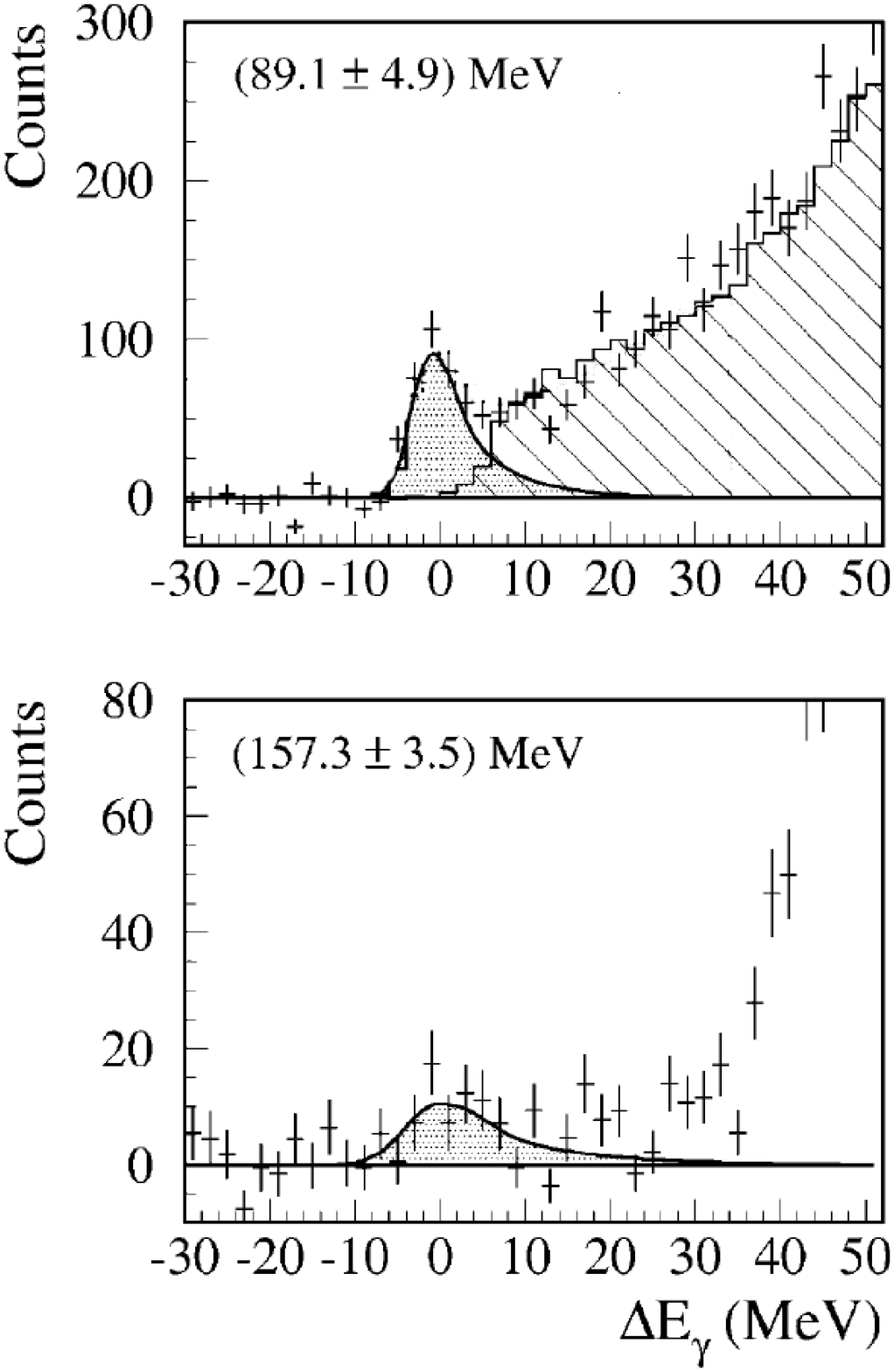}\includegraphics[width=5.3cm]{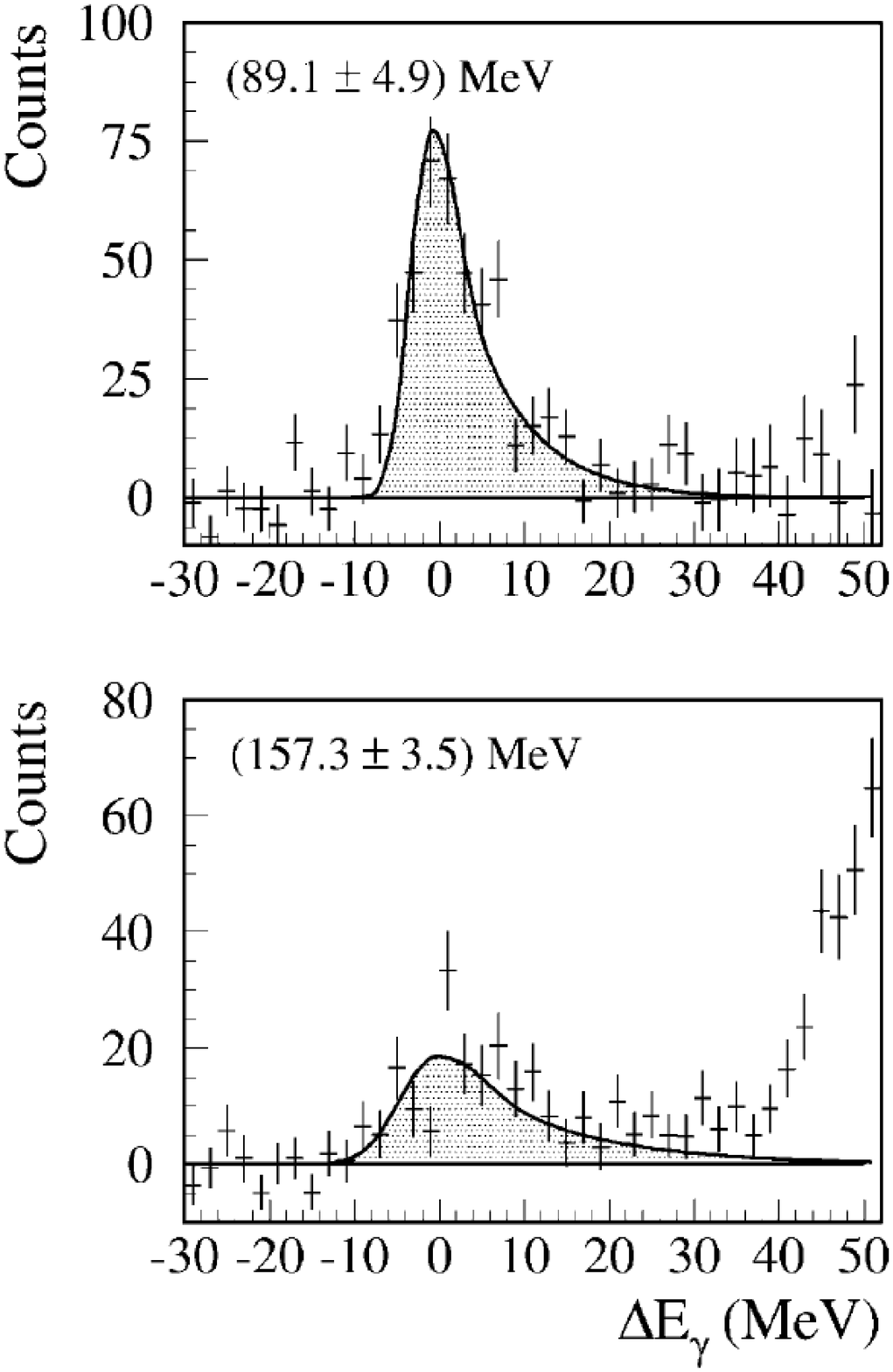}
\caption{\label{fig:TAPS_Emiss}TAPS event selection using missing energy.
Left panels show $\theta=59^{\circ}$, right $\theta=133^{\circ}$.
The shaded area shows the simulated response to elastically scattered photons.
The upper panels show the spectra below the pion threshold (at incoming $E_{\gamma}=89.1$ MeV) with hatching
to show the measured charged particle distribution scaled to the spectrum.
The lower panels are above the pion threshold (at incoming $E_{\gamma}=157.3$ MeV) and therefore exhibit
a rise at the higher side of the missing energy spectrum \cite{OlmosdeLeon:2001zn}.
 }
\end{figure}

\begin{figure}
\centering\includegraphics[width=6cm]{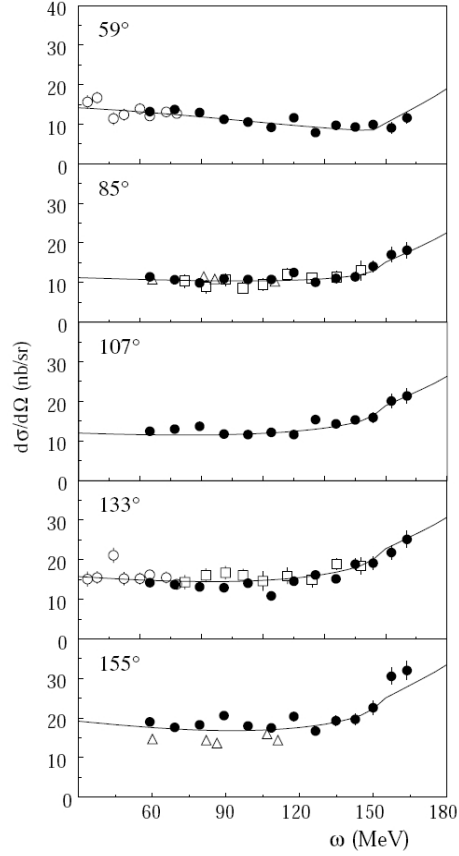}
\caption{\label{fig:TAPS_xsec}The TAPS differential cross sections \cite{OlmosdeLeon:2001zn} (full circles).
Additional data from \cite{Baranov:1974ec} (triangles), \cite{Federspiel:1991yd} (open circles) and \cite{MacGibbon:1995in} (squares),
with Dispersion Relation curves using the $\pi$-production multipoles
of \cite{Arndt:1995ak}. }
\end{figure}

\begin{figure}
\centering\includegraphics[width=8cm]{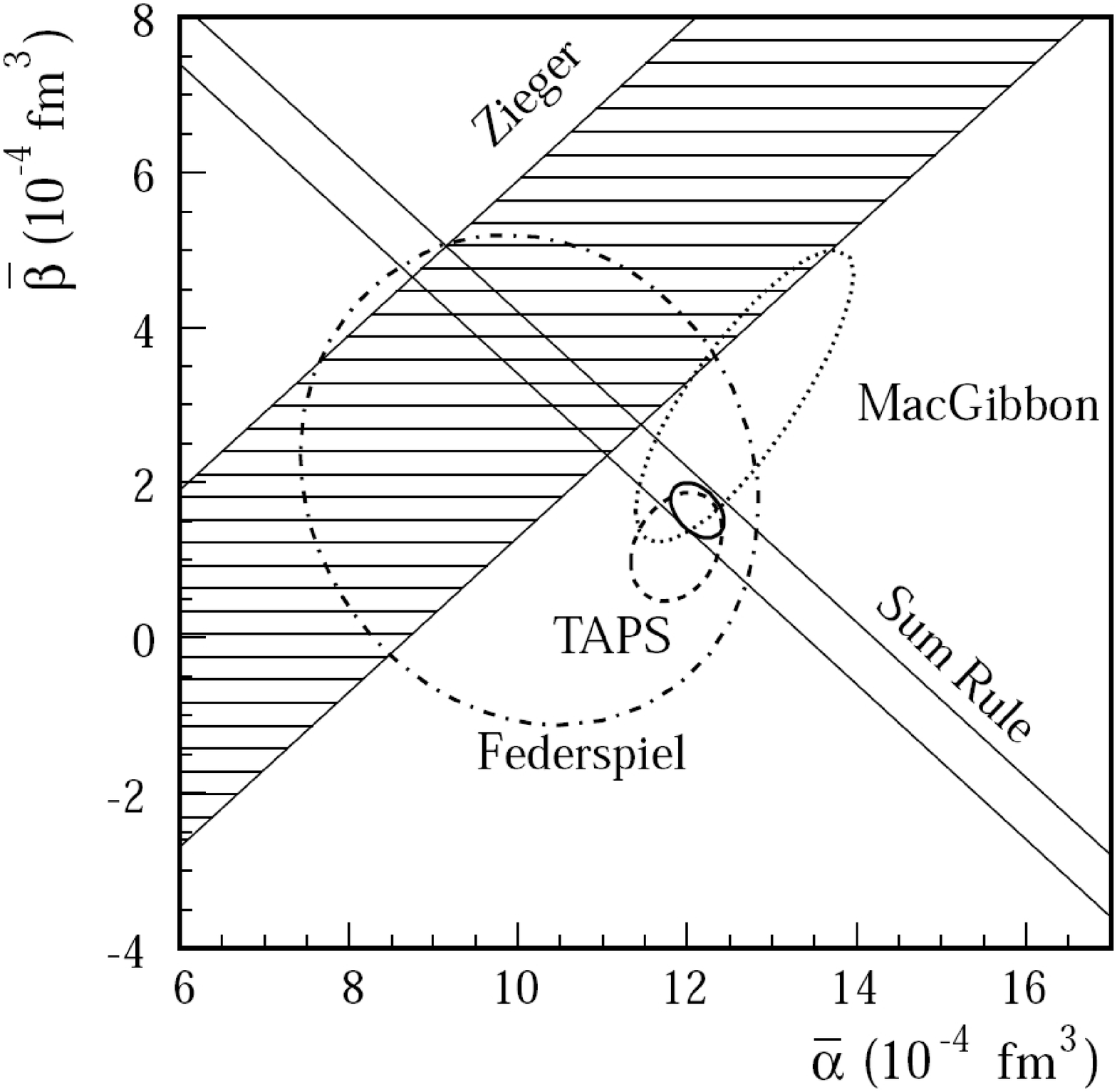}
\caption{The $\bar{\alpha}$ and $\bar{\beta}$ results in RCS, from \cite{Federspiel:1991yd,MacGibbon:1995in,OlmosdeLeon:2001zn,Zieger:1992jq}.
The thick solid line shows the result of the global fit in \cite{OlmosdeLeon:2001zn}. 
}
\label{Flo:alpha_beta} 
\end{figure}

In 2001 Olmos de Leon {\it et al.\/} reported on one such RCS measurement in the
Tagged Photon Facility at MAMI \cite{OlmosdeLeon:2001zn}. Here they
scattered energy-tagged Bremsstrahlung photons with $55<E_{\gamma}<165\, MeV$
from the Glasgow Photon Tagger on a $20\, cm$ liquid Hydrogen target
to measure the differential cross section of Compton scattering and
thereby extract $\ale$ and $\bem$. The scattered photons
were detected in the TAPS Barium Fluoride calorimeter which was arranged
in six blocks as shown in figure \ref{fig:TAPS_Setup}. The Compton
events were identified on the basis of the scattered photon missing
energy: the difference between the measured photon energy and that
calculated from the incoming beam energy and outgoing photon angle,
assuming Compton kinematics (figure \ref{fig:TAPS_Emiss}). The resulting
differential cross sections provided improved statistical accuracy
and covered a larger kinematical range than the existing data at that
time (figure \ref{fig:TAPS_xsec}). The extraction of $\ale$
and $\bem$ was performed using Dispersion Relation calculations
from L'vov \cite{L'vov:1996xd} and provided a very significant improvement
in the accuracy of such measurements as can be seen in figure \ref{Flo:alpha_beta}.
With this measurement and the many analyses that followed \cite{Schumacher:2005an},
the values of $\ale$ and $\bem$ became reasonably well
accepted.

The measurement of RCS is a complicated process due to its very low
cross section, of the order of tens of nanobarns. Below the pion production
threshold, electromagnetic events (pair production in the beam line
collimation and target) must be suppressed as far as possible. In
the case of the measurement described above, this was achieved by
the use of plastic veto detectors. As the veto suppression was not
absolute, events with veto hits were used to model the background
that escaped veto recognition. These were then scaled to the shape
of the measured data and subtracted from the region of interest in
the missing energy distributions (figure \ref{fig:TAPS_Emiss}). Above the pion production threshold,
things become even more difficult as it is relatively easy for one
of the two $\pi^{0}$ decay photons to go undetected, leaving a single
scattered photon in the detector. As the $\pi^{0}$ production cross section
is two orders of magnitude larger than that for Compton Scattering, this 
can be a very problematic source of contamination. This contribution was removed from
the above MAMI results by simulation of the scattered photon contribution
and subtraction from the missing energy spectrum.

\subsection{The spin polarisabilities}

When one goes to higher beam energies, the expansion given in 
eq.(\ref{Eqn:RCS}) has to be extended to the next order ($O(\omega^{3})$),
at which point four new terms, the spin polarisabilities, enter. These
parameterise the spin response of the nucleon to a changing electromagnetic
field and are known as $\gamma_{E1E1}$, $\gamma_{M1M1}$, $\gamma_{E1M2}$
and $\gamma_{M1E2}$ , where the subscript refers to the incoming
and outgoing photon multipoles. In order to extract these individual
spin polarisabilities, one has to use doubly polarised observables. It is
however possible to extract two linear combinations of the four variables 
without resorting to these lengths:
$\gamma_{0}$ can be extracted through measurement of the GDH
sum rule with a singly polarised measurement (yielding 
$\gamma_{0}=(-1.87\pm0.08_{stat}\pm0.10_{syst}) \cdot 10^{-4}fm^{4}$ 
\cite{Ahrens:2001qt}), and $\gamma_{\pi}$ can be extracted at the backward
limit of the unpolarised Compton scattering cross section. These can be expressed
in terms of the four spin polarisabilities such that
%
%
\begin{eqnarray*}
\begin{array}{lll}
\gamma_{0} & = & -\gamma_{E1E1}-\gamma_{M1M1}-\gamma_{E1M2}-\gamma_{M1E2}   \\
\gamma_{\pi} & = & -\gamma_{E1E1}+\gamma_{M1M1}-\gamma_{E1M2}+\gamma_{M1E2} \, . \\
\end{array}
\end{eqnarray*}
%
%
In the MAMI measurement  \cite{OlmosdeLeon:2001zn}, 
a new extraction of $\gamma_{\pi}$ was
performed, giving a value of 
$\gamma_{\pi}=(-35.9\pm2.3_{stat}\mp0.4_{syst}) \cdot 10^{-4}fm^{4} $,
without sum rule constraint and 
$\gamma_{\pi}=(-36.1\pm2.1_{stat}\mp0.4_{syst}\pm0.8_{mod.}) \cdot 10^{-4}fm^{4} $  
with a constraint from the Baldin Sum Rule ($\bar{\alpha}+\bar{\beta}=(13.8\pm0.4) \cdot  10^{-4}fm^{3}$)
as derived in the TAPS paper \cite{OlmosdeLeon:2001zn} from SAID,
MAMI data and \cite{Babusci:1997ij}.

The TAPS measurement had therefore achieved a new level of accuracy
in the measurement of $\ale$ and $\bem$ and produced
new values for the Baldin Sum Rule and $\gamma_{\pi}$. 
However it, as in the other Compton Scattering experiments 
of its generation,
was not able to access the spin polarisabilities individually. 
To achieve this, one requires
to perform single and double spin asymmetry measurements of RCS. The
measurement of simple differential cross sections had already proved
challenging due to the high level of background involved in the accurate
measurement of such a low cross section process. The current measurement
program at MAMI, however, aims to meet this challenge, and measure
three spin asymmetries in RCS between photon energies of $200<E_{\gamma}<300\, MeV$
\cite{Hornidge:2009}. To achieve this, one requires a polarised proton
target and circularly and linearly polarised photon beams.

In order to measure with a polarised proton target, one can no longer
use pure Hydrogen as, due to its diatomic nature, it is unpolarisable.
Instead, one has to polarise other materials such as frozen Butanol,
which have been radical doped to allow their polarisation through
the application of high magnetic fields, ultra low temperatures and
microwave pumping in the process known as Dynamic Nuclear Polarisation
(DNP). For the study of RCS on the proton, this introduces a whole
range of coherent and quasi-free background processes on the heavy
nuclei within the target material itself.

\begin{figure}
\centering\includegraphics[width=9cm]{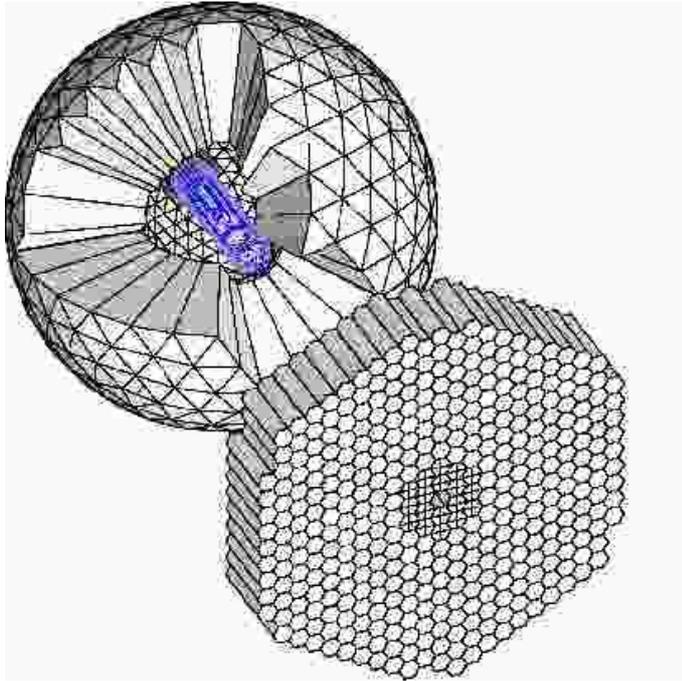}
\caption{\label{fig:CB_setup}The CB and TAPS detector system. The 
beam enters
from the upper left, scattering in the target at the centre of the
Crystal Ball. Particles traveling between $21^{\circ}$ and $159^{\circ}$
in $\theta_{lab}$ travel through the Particle Identification Detector
and two Multi-Wire Proportional Chambers then into the 672-element
NaI Crystal Ball. Particles traveling in $\theta_{lab}<21^{\circ}$
impinge on the TAPS detector system with plastic veto counters in
front of a spectrometer composed of BaF$_{\mbox{2}}$ with two highly
segmented PbWO$_{\mbox{4}}$ inner rings for enhanced rate capability.}
\end{figure}

In order to overcome this problem, one can take advantage of the hermetic
nature of the Crystal Ball (CB)
detector setup. This consists of a spectrometer
setup covering 97\% of $4\pi$, with the Crystal Ball surrounding
the target and the TAPS spectrometer in a forward wall configuration
(figure \ref{fig:CB_setup}). Particle identification and charged
particle tracking in the CB are provided by the Particle Identification
Detector (a barrel of 24 plastic scintillators), and two Multi-Wire
Proportional Chambers (respectively). In TAPS, one can use time-of-flight,
pulse shape analysis and the plastic veto detectors in order to separate
out particle species. With this large acceptance system, it is possible
to require both the proton and the photon and thereby separate the
coherent processes on heavier nuclei from the quasi-free contributions
and the RCS on the proton. In addition to this, by measurement on
a carbon target with the same number of nucleons as all the heavy
nuclei in the Butanol target (Oxygen, Carbon and the He cooling mixture),
one can obtain a background {}``quasi-free'' sample that also matches
the other target background issues (entry and exit foils, Butanol
container etc.). These data are then directly scaled and subtracted.
In the end, one is left with the processes which occur on the proton
itself, RCS and $\pi^{0}$ production where one of the photons has
scattered out of the detector acceptance. As with the TAPS measurement,
the scattered photon $\pi^{0}$ contribution is simulated, scaled
to the measured data and then subtracted.

\subsection{Current experimental status and future outlook}

This singly- and doubly-polarised experimental program is already
underway at MAMI  
and clear RCS signals from the proton have been extracted
from the data. The entire program seeks to measure three asymmetries: $\Sigma_{3}=\frac{\sigma^{\parallel}-\sigma^{\perp}}{\sigma^{\parallel}+\sigma^{\perp}}$,
with an unpolarised Hydrogen target and linearly polarised photon
beam; $\Sigma_{2z}=\frac{\sigma_{+z}^{R}-\sigma_{+z}^{L}}{\sigma_{+z}^{R}+\sigma_{+z}^{L}}=\frac{\sigma_{+z}^{R}-\sigma_{-z}^{R}}{\sigma_{+z}^{R}+\sigma_{-z}^{R}}$,
with a circularly polarised photon beam and longitudinally polarised
target; and $\Sigma_{2x}=\frac{\sigma_{+x}^{R}-\sigma_{+x}^{L}}{\sigma_{+x}^{R}+\sigma_{+x}^{L}}=\frac{\sigma_{+x}^{R}-\sigma_{-x}^{R}}{\sigma_{+x}^{R}+\sigma_{-x}^{R}}$
with a circularly polarised photon beam and transversely polarised
target. 

To extract the polarisabilities from the data it is necessary
to employ some theoretical model. In order to limit the model dependence
of the resulting polarisability values, we will be investigating the
data with several different models: Dispersion Relations \cite{Pasquini:2007hf},
a unitary and causal effective field theory based on the chiral Lagrangian
\cite{Gasparyan:2011yw} and chiral effective field theory \cite{Hildebrandt:2003md},
amongst others. The three asymmetry measurements will then be used
in a combined fit to extract experimental values for the often predicted,
but never yet experimentally accessed spin polarisabilities of the
proton.

MAMI is not the only laboratory working on the polarisability challenge,
through the measurement of Compton Scattering. There is also an active
polarisability program at the High Intensity Gamma Source (HIGS) at
the Triangle Universities' Nuclear Laboratory (TUNL) \cite{Miskimen:2009}.
The HIGS photon beam comes from 
the laser light produced by a free electron laser backscattering from the electrons in a storage ring, 
rather than Bremsstrahlung. 
While MAMI can cover a wide energy range with one measurement \cite{tagger},
HIGS measures at one precisely known energy at a time, but with 
100\% linearly or circularly polarized photons.
HIGS covers a
lower energy range than MAMI, 
currently with beam energies of up to 100 MeV, but with planned upgrades to reach 160 MeV. 
In this region, HIGS can make very accurate measurements of the
beam- and beam-target-asymmetries using an active polarised target
and the HINDA detector array - an arrangement of large volume NaI
detectors to give a very precise photon energy resolution, with 
active shields to veto cosmic ray backgrounds and energy leakage from the central NaI detector.
At these energies, the
recoiling nucleon does not escape the target and the use of an active
target to detect the recoiling nucleon in-situ is essential to make
a clean measurement. 

The HIGS measurement will have some sensitivity
to the spin polarisabilities of the nucleon, and should start to provide
high quality data almost as soon as the active polarised target is ready.
Sensitivity to the spin polarisabilities is highest in the 
Delta resonance
region where MAMI will measure. However, as one relies on theory to
extract the polarisabilities from the resulting asymmetry measurements,
the constraint of that theory through high quality data over a wide
kinematical range reduces the model dependence substantially. Thus the
combined measurements of both HIGS and MAMI should provide a first
experimental insight into the as yet unmeasured depths of the spin
polarisabilities of the nucleon.


\section{Virtual Compton Scattering} \label{sec-vcs}

In Virtual Compton Scattering (VCS) the \ $\gamma^{*} N \to \gamma N$ \  process is probed by a virtual, spacelike photon of four-momentum transfer squared $Q^2$. The polarisabilities become $Q^2$-dependent observables called generalized polarisabilities (GPs). They describe the spatial variation of the polarisation response of the nucleon, 
since $Q^2$ is related to the distance scale (by Fourier transform).

These new observables have been measured on the proton by electron scattering experiments, via exclusive photon electroproduction: \ $\epg $
\ in the low-energy regime. This kinematical domain means that the total energy $\sqrt{s}$, or  $W$ of the $(\gamma p)$ system in its center-of-mass is typically of the order of one nucleon plus one pion mass. The \ $\epg $ \ process at low energy has been the subject of several dedicated experiments, at MAMI~\cite{Roche:2000ng,Bensafa:2006wr,Janssens:2008qe}, the Thomas Jefferson National Accelerator Facility (JLab)~\cite{Laveissiere:2004nf,Laveissiere:2008zn} and MIT-Bates~\cite{Bourgeois:2006js}.
The observables that one can measure are GPs and structure functions.
For details, we refer the reader to existing reviews~\cite{Guichon:1998xv,d'Hose:2000xr,HydeWright:2004gh,Fonvieille:2004rb,d'Hose:2006xz}.

\subsection{Generalized polarisabilities}\label{subsec-gp}

%
The formalism of VCS on the nucleon was initially explored in~\cite{BergLindner:1961} and the concept of generalized polarisabilities was first introduced in~\cite{DrechselArenhovel:1974}. The first application of the  Low Energy Theorem (LET) to VCS was established in~\cite{Guichon:1995pu}, opening a new field of investigation in low-energy hadron physics.

\begin{figure}[htb]
\centering\includegraphics[width=8cm]{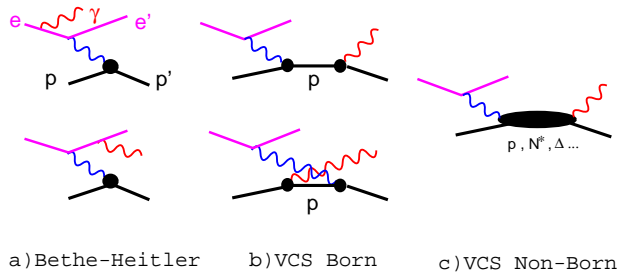}    
\caption{Feynman graphs of photon electroproduction. In the VCS Born graphs the intermediate state is a proton on-mass-shell. The $\pi^0$ exchange in the $t$-channel is included in the Non-Born part.}
\label{fig-bh+b+nb}
\end{figure}

As in RCS, the LET allows the separation of the structure-independent and structure-dependent parts of the Compton amplitude, called respectively Born and Non-Born (figures.~\ref{fig-bh+b+nb}-b, ~\ref{fig-bh+b+nb}-c). The GPs are obtained from the multipole decomposition of the Non-Born amplitude, taken in the limit $q'_{cm} \to 0$, where $q'_{cm}$ is the modulus of the momentum of the final real photon in the $\gamma p$ center-of-mass (noted CM hereafter). In this ``zero-frequency'' limit, the final photon is analogous to a static field, and 
%
%
%
 {\it ``VCS at threshold can be interpreted as 
%
%
%
electron scattering by a target which is in constant electric and magnetic fields''}~\cite{Guichon:1998xv}. The GPs have thus some similarity in nature with form factors; they describe the internal distribution of electric charge, magnetization and spin of a deformed nucleon.

%
As in RCS, the GPs depend on the quantum numbers of the two electromagnetic transitions involved in the Compton process, and usually a multipole notation is adopted. Initially ten independent lowest-order GPs were defined~\cite{Guichon:1995pu}. It was shown~\cite{Drechsel:1998xv,Drechsel:1998zm} that nucleon crossing and charge conjugation symmetry  reduce this number to six: two scalar (S=0) and four spin, or spin-flip, or vector GPs (S=1). They can be conveniently defined as shown in table~\ref{tab-dipole-gps}. We note that the GPs are functions of $q_{cm}$, the virtual photon momentum in the CM, or equivalently the photon virtuality taken in the limit  $q'_{cm} \to 0$, which  will be denoted $Q^2$ for simplicity. The two scalar GPs, electric and magnetic, are defined as:
%
%
\begin{eqnarray*}
\begin{array}{lll}
\alpha_E (Q^2) & = & - P^{(L1,L1)0} (Q^2) \cdot ( {e^2 \over 4 \pi} \sqrt{3 \over 2} ) \\
\beta_M  (Q^2) & = & - P^{(M1,M1)0} (Q^2) \cdot ( {e^2 \over 4 \pi} \sqrt{3 \over 8} ) \\
\end{array}
\end{eqnarray*}
%
%
which coincide in the limit $Q^2 \to 0$ with the usual static RCS polarisabilities $\alpha_E$ and $\beta_M$. Some of the spin GPs  ($\gamma_3, \gamma_2 + \gamma_4$) also have a corresponding quantity in RCS \cite{Drechsel:1998zm}.

We mention here the fully covariant framework of ref.~\cite{L'vov:2001fz}, in which three scalar GPs are introduced instead of two. In particular two electric GPs, $\alpha_L$ and $\alpha_T$, are needed to fully reconstruct the spatial distribution of the electric polarisation.

\begin{table}[htbp]
\begin{center}
\caption{The standard choice for the six dipole GPs.
The original notation in column 1, $P^{(\rho' L' , \rho L)S}$, uses  the polarisation state  $\rho (\rho')$ of the initial (final) photon,  the angular momentum $L (L')$ of the transition, and the non spin-flip $(S=0)$ or spin-flip $(S=1)$  of the nucleon. The multipole notation in column 2 uses the electric, magnetic and longitudinal  $(E,L,M)$ multipoles. The six listed GPs correspond to the lowest possible order in $q'_{cm}$, i.e. a  dipole final transition $(l'=1)$.
Column 3 gives the correspondence in the RCS limit, defined by $Q^2 \to 0$ or $q_{cm} \to 0$.
}
\label{tab-dipole-gps}
\begin{tabular}{|c|c|c|}
\hline
\  $P^{(\rho ' L', \rho L ) S } (q_{cm})$ \ & \ $P^{(f,i)S} (q_{cm}) \ $ 
& RCS limit  \\
\hline
 $P^{(01,01)0}$ & $P^{(L1,L1)0}$ & 
$ - {4 \pi \over e^2} \sqrt{2 \over 3} \ \ale $   \\
 $P^{(11,11)0}$ & $P^{(M1,M1)0}$ & 
$ - {4 \pi \over e^2} \sqrt{8 \over 3} \ \bem $  \\
 $P^{(01,01)1}$ & $P^{(L1,L1)1}$ & 0 \\
 $P^{(11,11)1}$ & $P^{(M1,M1)1}$ & 0  \\
 $P^{(01,12)1}$ & $P^{(L1,M2)1}$ & 
$ - {4 \pi \over e^2} {\sqrt{2} \over 3} \ \gamma_3$  \\
 $P^{(11,02)1}$ & $P^{(M1,L2)1}$ & 
$ - {4 \pi \over e^2} { 2 \sqrt{2} \over 3 \sqrt{3}}  (\gamma_2 + \gamma_4)$ \\
\hline
\end{tabular}
\end{center}
\end{table}

\subsection{Theoretical models}\label{subsec-theo-models}


GPs are valuable observables to investigate nucleon structure. They have the potential  to shed new light on the interplay between nucleon-core excitations and pion-cloud effects. They have been calculated by a number of theoretical models dealing with these ingredients in various ways.
In heavy baryon chiral perturbation theory (HBChPT), the polarisabilities are pure one-loop effects to leading order in the chiral expansion~\cite{Bernard:1993bg}, emphasizing the role of the pion cloud. In this approach the scalar GPs have been calculated to order $p^3$~\cite{Hemmert:1997at,Hemmert:1996gr,Hemmert:1999pz}.
The spin GPs (and the scalar polarisabilities in RCS) have been calculated to order $p^4$~\cite{Kao:2002cn,Kao:2004us}. The first nucleon resonance $\Delta (1232)$ is taken into account either by local counterterms (ChPT,~\cite{Bernard:1993bg}) or as an explicit degree of freedom (small scale expansion SSE of~\cite{Hemmert:1999pz}).
In non-relativistic quark constituent models~\cite{Guichon:1995pu,Liu:1996xd,Pasquini:2000ue,Pasquini:2001pj} the GPs involve  the summed contribution of all nucleon resonances but do not incorporate a direct pionic effect. The calculation of the linear-$\sigma$ model~\cite{Metz:1996fn,Metz:1997fr} involves all  fundamental symmetries but does not include the $\Delta$ resonance. The effective lagrangian model~\cite{Vanderhaeghen:1996iz} includes resonances and the pion cloud in a more phenomenological way. A calculation of the electric GP was made in the Skyrme model~\cite{Kim:1997hq}.
Lattice calculations are for the moment limited to polarisabilities in RCS~\cite{Detmold:2010ts}.

 The Low Energy Theorem~\cite{Guichon:1995pu,Guichon:1998xv} and the Dispersion Relation formalism developed for RCS and VCS~\cite{Pasquini:2001yy,Drechsel:2002ar}, are of special importance in the interface to experimental analyses and are detailed in the next section.


As seen in section \ref{sec-rcs}, the proton polarisabilities $\ale$ and $\bem$ are small quantities at the real photon point. The further smallness of  $\bem$ relative to $\ale$ is traditionally explained by the existence of two different contributions of opposite sign, para- and diamagnetic, which nearly cancel. 
Regarding the $Q^2$-dependence, most theoretical calculations (apart from quark models) find that $\beta_M$ first rises with $Q^2$ and then decreases. This is explained by the importance at small $Q^2$, or long-distance, of the diamagnetism due to the pion cloud. This contribution is partially cancelled and eventually dominated at high $Q^2$ by the short-distance paramagnetism of the quark core. The electric GP is usually predicted to fall off smoothly  with $Q^2$ like a dipole. A compilation of theoretical predictions for the six GPs can be found e.g. in figure 2 of ref.~\cite{Pasquini:2001pj}.

\subsection{Theoretical tools for VCS analyses}

The GPs do not enter the $\epg$ cross section in a straightforward way; a theoretical tool is needed to extract them from an experiment. Up to now, two approaches have been used for this purpose: the Low Energy Theorem (LET) ~\cite{Guichon:1995pu} and  Dispersion Relations ~\cite{Pasquini:2001yy}.

\subsubsection{The Low Energy Theorem and the VCS structure functions}\label{subsec-lex-theo}


The  $\epg$ reaction is the coherent sum of the VCS processes (figures \ref{fig-bh+b+nb}-b, \ref{fig-bh+b+nb}-c) and the Bethe-Heitler one (BH, figure \ref{fig-bh+b+nb}-a). The electroproduction amplitude is decomposed as: $T^{e p \gamma} = T^{BH} + T^{Born} + T^{NonBorn}$. In this sum the two first terms (BH+Born) are known and calculable, with the proton electromagnetic form factors $G_E^p$ and $G_M^p$ as inputs. The third amplitude $T^{NonBorn}$ contains the unknown GPs, and the LET specifies how to access these observables analytically.

According to the LET, or LEX (Low-energy EXpansion), the amplitude $T^{e p \gamma}$ is expanded in powers of $q'_{cm}$. As a result, the (unpolarised)  $\epg$ cross section at small $q'_{cm}$ can be written as:
\begin{eqnarray}
 d^5 \sigma &=&   
 d^5 \sigma ^{BH+Born}  \ + \ q'_{cm} \cdot \phi \cdot \Psi_0 \ + \ {\cal O}(q'^2_{cm}) \ \ \  
\label{eq-let1} 
\end{eqnarray} 
$\phi$ being a phase-space factor. The notation 
$d^5 \sigma$ stands for $d^5 \sigma / dk'_{elab} d \Omega'_{e lab} d \Omega_{cm}$ where $k'_{elab}$ is the scattered electron momentum, $d \Omega'_{e lab}$ its solid angle and  $d \Omega_{\gamma cm}$  the solid angle of the outgoing photon (or proton) in the CM. The $\Psi_0$ term comes from the interference between the Non-Born and the BH+Born amplitudes at lowest order in $q'_{cm}$; it gives the leading polarisability effect in the cross section. This approach is valid below the pion production threshold, i.e. as long as the Non-Born amplitude remains real. Eq.(\ref{eq-let1})  clearly states that information on the GPs is obtained by measuring the {\it deviation} from BH+Born.

\begin{figure}[htb]  
\centering\includegraphics[width=8.3cm]{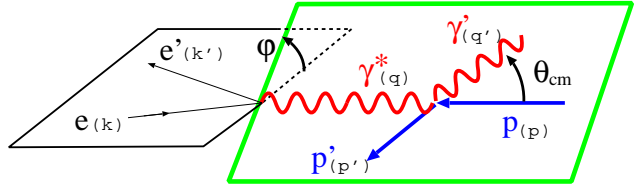}  
\caption{$(ep \to ep \gamma)$ kinematics and Compton angles $(\theta_{cm},\varphi)$ in the $\gamma p$ center-of-mass.
}
\label{fig-epgamma-kinem}
\end{figure}

The $\Psi_0$ term contains three structure functions (or VCS response functions) $P_{LL}$, $P_{TT}$ and $\plt$, under the form:
\begin{eqnarray}
\Psi_0 &=& v_1 \cdot 
(P_{LL} - {\displaystyle 1 \over \displaystyle \epsilon} P_{TT}) 
\ + \ v_2 \cdot  P_{LT}  
\label{eq-let2} 
\end{eqnarray} 
where $\epsilon$ is the usual virtual photon polarisation parameter and $v_1, v_2$ are kinematical coefficients depending on $(q_{cm},\epsilon,\theta_{cm},\varphi)$. $\theta_{cm}$ and $\varphi$ are the polar and azimuthal angles of the Compton scattering process in the CM (see figure~\ref{fig-epgamma-kinem}). The full expression of $v_1,v_2 $  can be found in ref~\cite{Guichon:1998xv}, as well as the expression of the structure functions versus the GPs. In particular, one has:
\begin{eqnarray}
\begin{array}{lll}
P_{LL}  & = &   
{ 4 \nucleonmass  \over \alpha_{em} } \cdot 
G_E^p(Q^2)\cdot  \alpha_E(Q^2)   \\
P_{TT}  & = & [P_{TTspin}]     \\
P_{LT}  & = & - { 2 \nucleonmass  \over \alpha_{em} } 
 \sqrt{ {q_{cm}^2 \over Q^2} } \cdot 
G_E^p(Q^2) \cdot \beta_M(Q^2) +  [P_{LTspin}]  \label{eq-sf12} 
\end{array}
\end{eqnarray} 
where $\alpha_{em}$ is the fine structure constant. The terms in brackets are the spin part of the structure functions, i.e. the following combinations of spin GPs:
\begin{eqnarray}
\begin{array}{lll}
P_{TT}  & = &  -3 G_M^p(Q^2) \, {q_{cm}^2 \over {\tilde q^0} } \cdot
(P^{(M1,M1)1}(Q^2) \\
\ & \ &  - {\sqrt{2}} \, {\tilde q^0} \cdot P^{(L1,M2)1}(Q^2)  \\
P_{LTspin}  & = & {3 \over 2} \, {  q_{cm} \sqrt{ Q^2} \over \tilde q^0 }  G_M^p(Q^2) \cdot P^{(L1,L1)1}(Q^2)  \label{eq-sfspin} 
\end{array}
\end{eqnarray} 
%
%
where $\tilde q^0$ is the CM energy of the virtual photon in the limit $q'_{cm} \to 0$.
The other terms in eq.(\ref{eq-sf12}) are the scalar parts of the structure functions. The important point is that $\pll$ is proportional to the electric GP, and the scalar part of $\plt$ is proportional to the magnetic GP. 

It is clear that using this LET approach, one cannot extract the six dipole GPs separately from an unpolarised experiment, since only three independent structure functions appear. At a given $Q^2$ and one single $\epsilon$,  one can determine only two quantities: $\plltt$ and $\plt$. As shown by eqs.(\ref{eq-let1}) and (\ref{eq-let2}), these unknowns can be determined experimentally by a linear fit of the difference $d^5 \sigma^{exp} - d^5 \sigma^{BH+Born}$, assuming the validity of the truncation to ${\cal O}(q'^2_{cm})$. This is the method used in LEX analyses. Then, to further isolate the scalar part in these structure functions, i.e. $\ale (Q^2)$ and $\bem (Q^2)$, a model input is required.

 A Low Energy Theorem has also been derived in the case of doubly polarised VCS~\cite{Vanderhaeghen:1997bx}. It allows in principle the disentanglement of all six lowest-order GPs, thereby accessing the spin GPs of the nucleon. Finally, we mention the recent work of ref.~\cite{Gorchtein:2009wz} in which the LET formalism for VCS has been established in the Breit frame instead of the CM frame, leading to six new dipole GPs instead of those of ref.~\cite{Guichon:1998xv}.

\subsubsection{Dispersion Relations} \label{subsec-dr-theo}

The Dispersion Relations (DR) formalism developed by B.Pasquini et al. \cite{Pasquini:2001yy,Drechsel:2002ar} for RCS and VCS offers an interesting alternative to extract structure functions and GPs from photon electroproduction experiments. An essential feature of the model is that it provides a rigorous treatment of the higher-order terms in the VCS amplitude, up to the $N \pi \pi$ threshold, by including resonances in the $\pi N$ channel. Therefore its validity extends to the region of the $\Delta(1232)$ resonance, where the LET does not hold.


The Compton tensor is parameterised through twelve invariant amplitudes $F_{i} (i=1,12)$. The GPs are expressed in terms of the non-Born part  $F_i^{NB}$ of these amplitudes at the point $t=-Q^2, \nu=(s-u)/4 \nucleonmass  =0$, where $s,t,u$ are the Mandelstam variables of the Compton scattering.
%
%
%
%
All of the $F_i^{NB}$ amplitudes, with the exception of two, fulfill unsubtracted dispersion relations. These $s$-channel dispersive integrals are calculated through unitarity. They are limited to the $\pi N$ intermediate states, which are considered to be the dominant contribution for describing VCS up to the $\Delta(1232)$ resonance region. 
In practice, the calculation uses the pion photo- and electroproduction multipoles given by MAID~\cite{Drechsel:1998hk}, in which both resonant and non-resonant production mechanisms are included. 

%
%
%
The amplitudes $F_1$ and $F_5$ have an unconstrained part beyond the $\pi N$ dispersive integral. Such a remainder is also considered for $F_2$. 
For $F_5$ this asymptotic contribution is dominated by the t-channel $\pi^0$ exchange, and with this input all four spin GPs are fixed.
For $F_1$ and $F_2$, an important feature is that in the limit $(t=-Q^2, \nu=0)$ their non-Born part is proportional to the GPs $\bem$ and $( \ale + \bem )$,  respectively. The remainder of $F _{1,2}^{NB}$ is estimated by an energy-independent function, noted $\Delta \beta$ and $\Delta (\alpha + \beta)$ respectively. This term  parameterises the asymptotic contribution and/or dispersive contributions beyond $\pi N$. For the magnetic GP one gets:
\begin{eqnarray}
\begin{array}{lll}
\bem (Q^2) & = & \beta^{\pi N}(Q^2) +  \Delta \beta \\
 \Delta \beta  & = & { \displaystyle [ \beta^{exp}   -  
\beta^{\pi N} ]_{Q^2=0} 
\over
\displaystyle ( 1 + Q^2/ \lamb^2 )^2 } \ .
\label{eq-dr-beta-0}
\end{array}
\end{eqnarray}
The sum $( \ale + \bem )$ follows a similar decomposition, and thus the electric GP too:
\begin{eqnarray}
\begin{array}{lll}
\ale (Q^2) & =  &  \alpha^{\pi N}(Q^2) +  \Delta \alpha \\ 
\Delta \alpha & = & { \displaystyle [ \alpha^{exp}   -  
\alpha^{\pi N} ]_{Q^2=0} 
\over
\displaystyle ( 1 + Q^2/ \lama^2 )^2 } \ .
\label{eq-dr-alpha-0}
\end{array}
\end{eqnarray}


In other words, the two scalar GPs are not fixed by the model; their unconstrained part is parameterised by a dipole form, as given by eqs.(~\ref{eq-dr-beta-0},~\ref{eq-dr-alpha-0}). This dipole form is arbitrary: the mass parameters $\lama$ and  $\lamb$ only play the role of intermediate quantities in order to extract VCS observables, with minimal model-dependence. In the DR calculation  $\lama$ and $\lamb$ are treated as free parameters, which can furthermore vary with $Q^2$. Their value can be adjusted by a fit to the experimental cross section, separately at each $Q^2$. Then the model is fully constrained and provides all VCS observables  at this $Q^2$: the scalar GPs as well as the structure functions, in particular $\plltt$ and $\plt$.


\subsection{VCS experiments} \label{sec-vcs-experiments}

\begin{figure}[htb]
\centering\includegraphics[width=8cm]{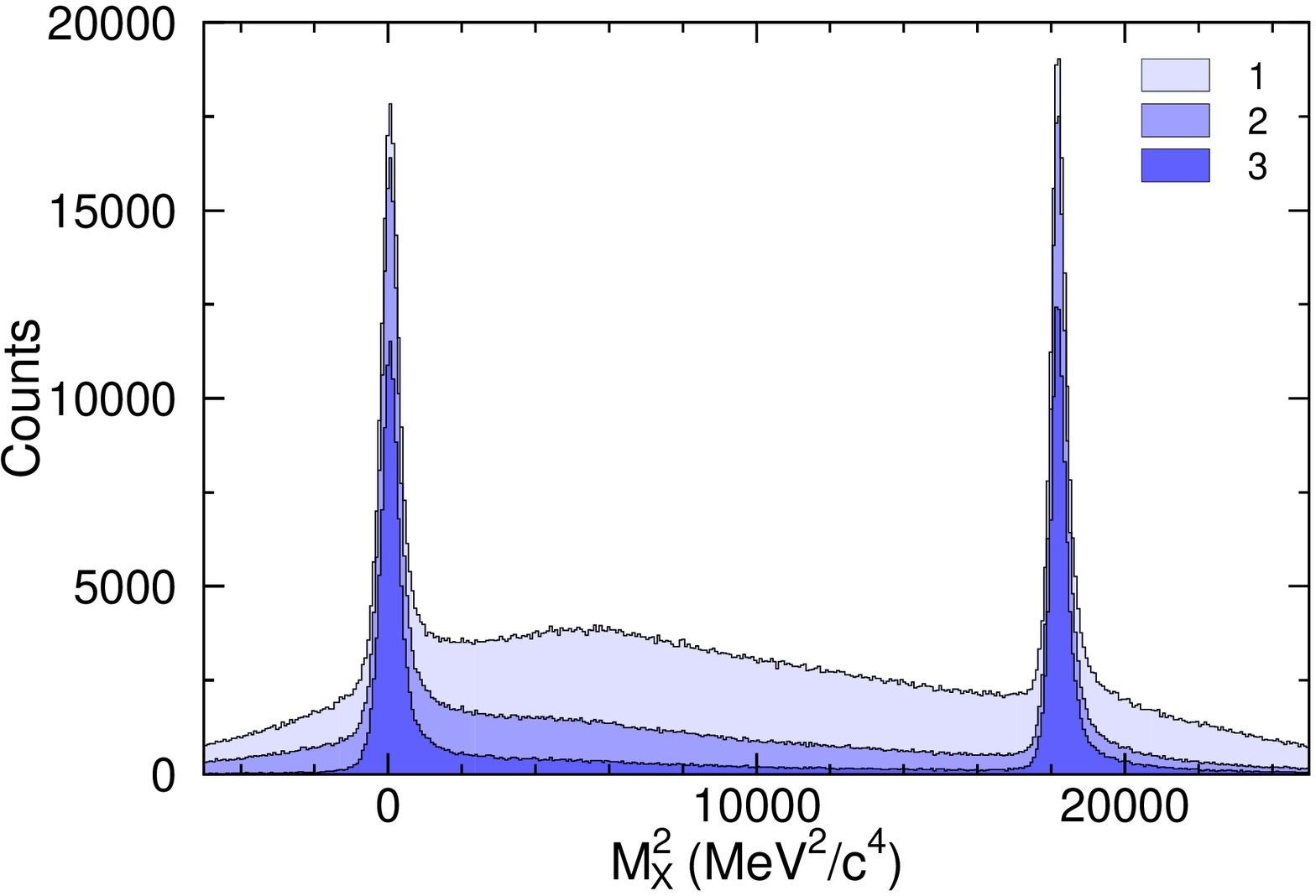}
\centering\includegraphics[width=8cm]{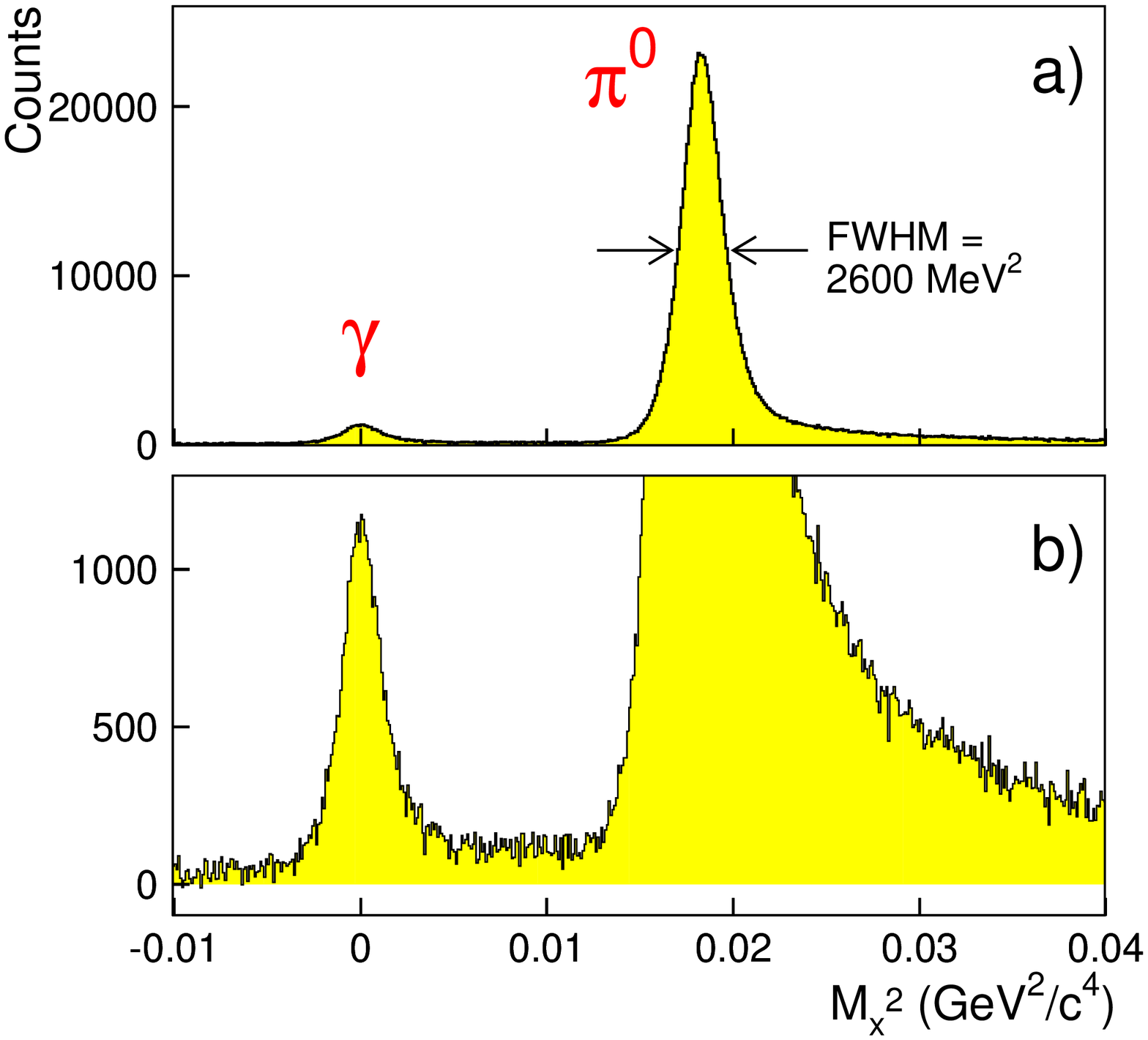}
\caption{\label{fig-vcs-missingmass}
The missing mass squared in the reaction (H(e,e'p)X), as obtained in two VCS experiments  at MAMI. Top: below the pion threshold (from~\cite{Janssens:2008qe}). Bottom: in the region of the $\Delta(1232)$ resonance (from~\cite{Bensafa:2006wr}) (histogram b) is histogram a) with a zoom in ordinate). 
In each case the $\gamma$ and $\pi^0$ peaks are extremely well separated.
}
\end{figure}

The photon electroproduction cross section is small (three electromagnetic vertices are involved) and thus requires high-performance equipment in order to be measured accurately. All VCS experiments are designed along the same lines, detecting the two charged final state (H(e,e'p)X) particles in coincidence and ensuring the reaction exclusivity by a missing mass technique. A good resolution in missing mass is therefore mandatory, and is usually reached (see figure~\ref{fig-vcs-missingmass}) through the use of small solid angle magnetic spectrometers.
As a consequence, VCS experiments require high luminosities, of the order of 10$^{37}$-10$^{38}$ cm$^{-2}$.s$^{-1}$, i.e. an intense electron beam and a liquid hydrogen target.

In the region of validity of the LET, i.e. below the pion threshold, the effect of the GPs in the $\epg$ cross section is small, about 10-15\%; furthermore to extract the GPs one needs a lever arm in the Compton angles, $\theta_{cm}$ and/or $\varphi$. Therefore the main difficulty of the VCS experiments is to obtain an accurate measurement of the absolute $(\epg )$ differential cross section, in a wide enough angular range, with systematic errors reduced to the few percent level.

\begin{figure}[htb]
\centering\includegraphics[width=9cm]{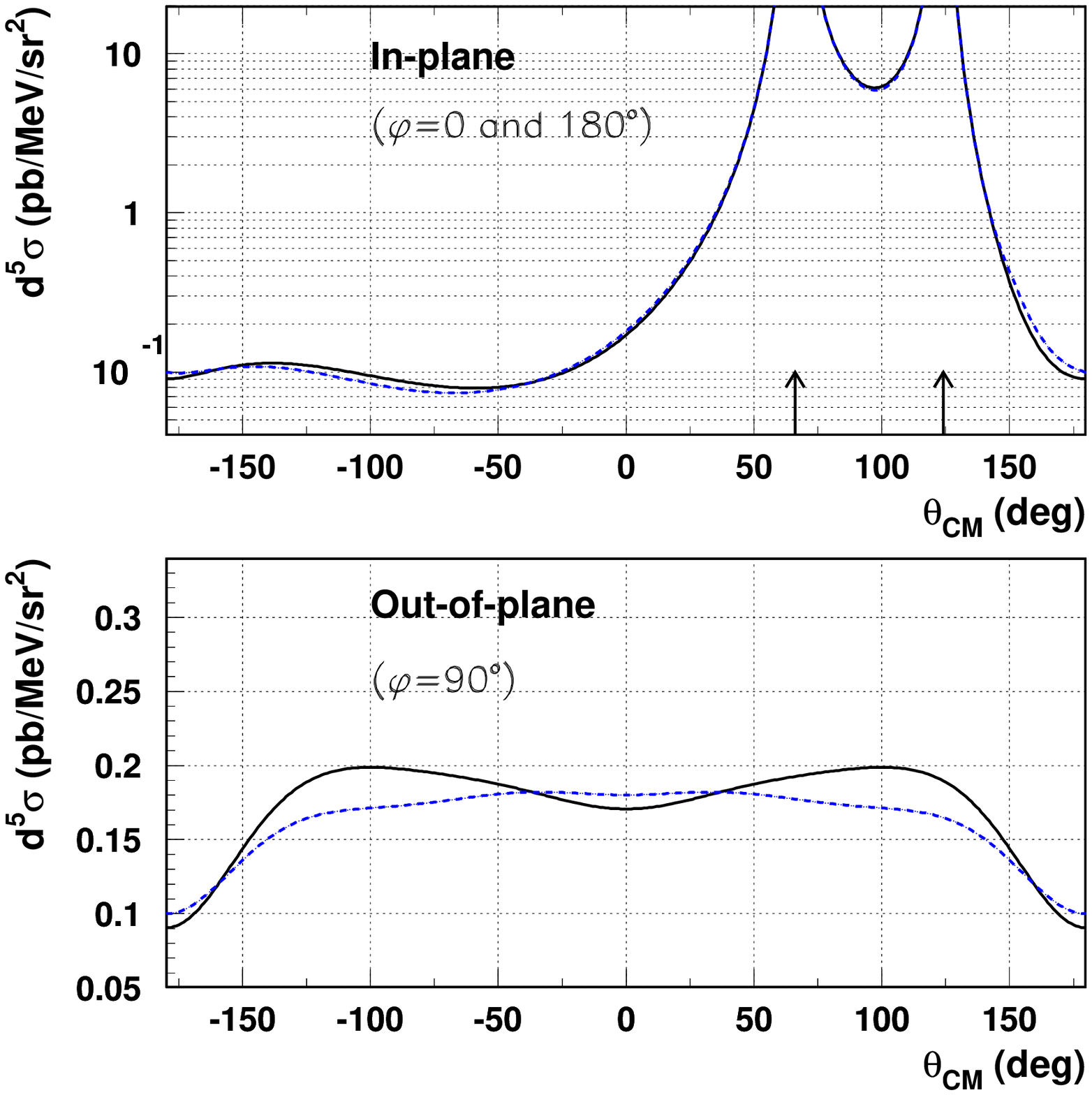}
\caption{\label{fig-vcs-theo-c-s}
The theoretical $\theta_{cm}$-behaviour of the photon electroproduction cross section, in-plane (top) and out-of-plane at $\varphi=90^{\circ}$ (bottom). In the top plot the BH peaks are indicated by arrows. We use the convention that $\theta_{cm}$ is negative in the hemisphere opposite to the peaks. The solid curve is the BH+Born cross section; the dashed curve includes a first-order GP effect as given by eq.(\ref{eq-let1}).
}
\end{figure}

The first experiment, which was performed at MAMI, focused on  in-plane kinematics; there the cross section exhibits two strong peaks due to the Bethe-Heitler radiation  (see figure~\ref{fig-vcs-theo-c-s}-top) and the GP-sensitive region lies in the hemisphere opposite to these peaks (negative $\theta_{cm}$ on the figure). The out-of-plane phase space was explored in later experiments, either with OOP spectrometers (Bates) or by exploiting the Lorentz boost from CM to Lab which focuses the emitted proton in a cone around the virtual photon direction (JLab). The out-of-plane cross section has a much smoother behaviour than in-plane; see figure~\ref{fig-vcs-theo-c-s}-lower panel.


To determine the experimental cross section  $d^5 \sigma$ one needs to calculate a five-fold solid angle. This is usually performed by a Monte-Carlo simulation, which has to reproduce the experiment faithfully, including the detector resolution, and have a realistic cross section (BH+Born) for the input event generator. The radiative corrections are also folded in, based on the formalism of ref.~\cite{Vanderhaeghen:2000ws}. The data are binned in $q'_{cm}$, because the polarisability effect in the cross section scales with $q'_{cm}$. The data are also binned in $\theta_{cm}$ and/or $\varphi$; this allows the performance of a meaningful fit of the structure functions $\plltt$ and $\plt$, based on the known angular dependence of the $v_1$ and $v_2$ coefficients of eq.(\ref{eq-let1}). Finally, an important point is the choice of proton form factors $G_E^p$ and $G_M^p$ in the analyses. Indeed, these quantities enter the calculation of the BH+Born cross section, and the GP effect is observed through a deviation from this cross section. Therefore the GP result is sensitive to the choice made for the form factors. A quantitative example of such a dependence can be found in tables 4 and 5 of ref.~\cite{Janssens:2008qe}.


\subsection{Results on the VCS structure functions} 
\label{sec-results-sf}

\begin{figure}[htb]
\centering\includegraphics[width=10cm]{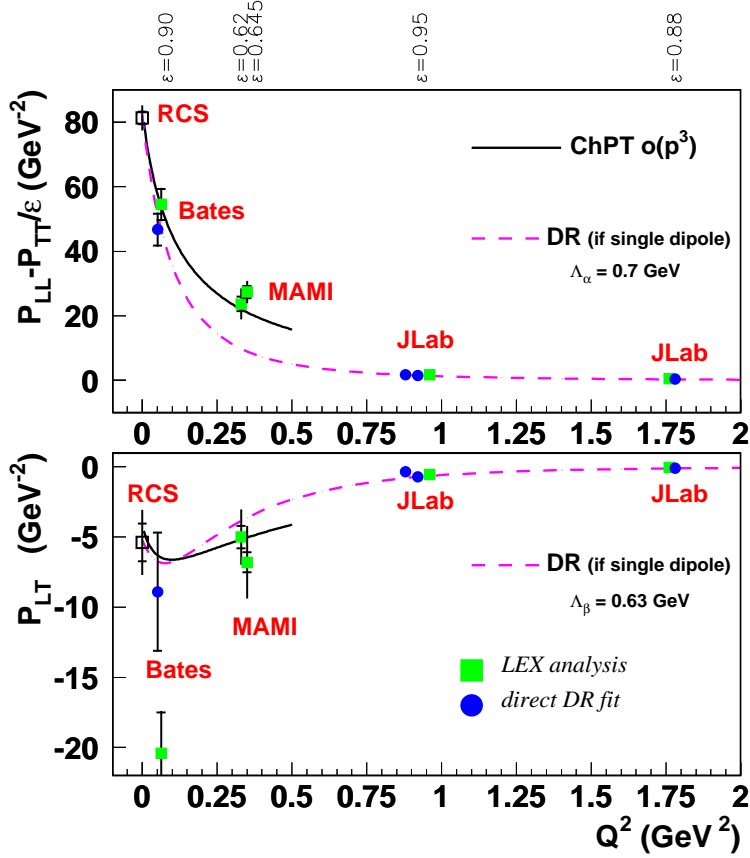}
\caption{\label{fig-sf}
World data \cite{Bourgeois:2006js,Roche:2000ng,Janssens:2008qe,Laveissiere:2004nf} on the structure functions  $\plltt$ (top) and $\plt$ (bottom) deduced from VCS analyses. LEX and DR results are indicated by squares and filled circles respectively. The point corresponding to RCS~\cite{OlmosdeLeon:2001zn} is also included. Some points are slightly shifted in abscissa  for visibility.  The inner (outer) error bars are statistical (total). The dashed curve shows a DR calculation with fixed parameters $\lama$ and $\lamb$ while the solid curve is the ChPT calculation at order $p^3$ \cite{Hemmert:1999pz}. $\epsilon=0.9$ is chosen to draw the theoretical curves  for $\plltt$.
}
\end{figure}

The structure functions  $\plltt$ and $\plt$ are determined by a fit to the cross-section data, as explained in sections \ref{subsec-lex-theo} and \ref{subsec-dr-theo}. The results are displayed in figure~\ref{fig-sf}. The lowest $Q^2$ is covered by Bates~\cite{Bourgeois:2006js}, the intermediate $Q^2$ by MAMI~\cite{Roche:2000ng,Janssens:2008qe} and the high $Q^2$ by JLab~\cite{Laveissiere:2004nf}. DR and LEX extractions (filled circles and squares in the figure) are in rather good agreement at high $Q^2$. At intermediate $Q^2$ this comparison still remains to be done consistently. At the lowest $Q^2$ (Bates experiment) a discrepancy was found for $\plt$, due to a breakdown of the LEX 
%
%
%
truncation 
%
%
%
in some of the chosen kinematics.

The data follow the global trend of the models, i.e. a more or less continuous fall-off for $\plltt$ and for $\plt$ a rather flat behaviour in the low-$Q^2$ region followed by a decrease to zero. These behaviours reflect in a large part the ones of the electric and magnetic GPs (see next section).

The HBChPT calculation (solid curve) at order $p^3$ \cite{Hemmert:1999pz} agrees well with the data at low $Q^2$. However, it should be kept in mind that the convergence of this calculation w.r.t. the chiral order is not reached, at least for the spin part of the structure functions~\cite{Kao:2002cn,Kao:2004us}. The DR calculation (dashed curve) drawn assuming a single dipole for $\Delta \alpha$ and $\Delta \beta$ (cf. section~\ref{subsec-dr-theo}), shows that this assumption is too simple and cannot account for $\plltt$ in its full measured range.

\subsection{Results on the electric and magnetic GPs} 
\label{sec-results-gp}

The difference between the two methodologies should be emphasised here: using the DR approach, the scalar GPs are extracted in a straightforward way from the experimental fit, whereas this is not the case for the LEX fit. In this latter case the spin parts, $\ptt$ and $P_{LT spin}$ have first to be subtracted from the measured  $\plltt$ and $\plt$. This subtraction has to be done using a model, since the spin GPs are not measured yet. To this aim one usually takes the DR model, so the whole GP picture becomes (DR-)model-dependent.

\begin{figure}[htb]
\centering\includegraphics[width=9cm]{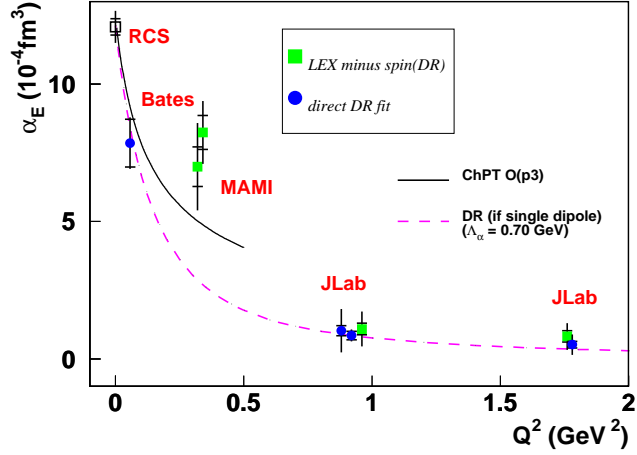}
\caption{\label{fig-world-alphae}
World data on the electric GP of the proton, with statistical and total error bar. The points are obtained by LEX (squares) or DR (filled circles) analyses. The theoretical curves correspond to the same calculations as in figure~\ref{fig-sf}. }
\end{figure}

Figure \ref{fig-world-alphae} shows the values of the electric GP. In the DR calculation of $\ale$ (dashed curve) the asymptotic part, noted $\Delta \alpha$ in eq.(\ref{eq-dr-alpha-0}), is largely dominant. Therefore this curve, which is calculated here for a fixed value of the $\lama$ parameter, falls approximately like a dipole. Similarly to what is observed for $\plltt$, all data points for $\ale$ are compatible with this shape, except in the region of the MAMI measurements. This behaviour near  $Q^2=0.3$ GeV$^2$ is puzzling and unexpected, since all models predict a smooth fall-off. It could be an indication of a physical ``structure'' at intermediate $Q^2$; however this statement should be taken with caution. More studies can and should be conducted on the existing samples, e.g. one could perform a consistent ``direct DR fit'' of $\ale$ at $Q^2=0.33$ GeV$^2$ (MAMI data) \cite{hf-dr-inprogress}, or  study in more detail the influence of the choice of proton form factors. Regarding this last point, it is clear that the recent MAMI data for $G_E^p$ and $G_M^p$ \cite{Bernauer:2010wm} provides a new and precise parameterisation, which will improve the accuracy of the determination of VCS observables at intermediate $Q^2$.

\begin{figure}[htb]
\centering\includegraphics[width=9cm]{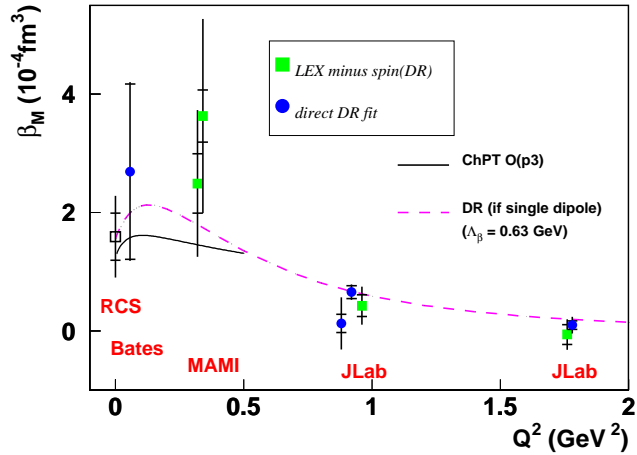}
\caption{\label{fig-world-betam}
World data on the magnetic GP of the proton, with statistical and total error bar. The points are obtained by LEX (squares) or DR (filled circles) analyses. See  previous figure for the theoretical curves.
 }
\end{figure}

 We mention a recent paper \cite{Gorchtein:2009qq} in which the $Q^2$-behaviour of the electric GP is described by a DR calculation, plus a gaussian contribution centered near $Q^2$=0.3 GeV$^2$ and introduced in order to agree with  the measured $\plltt$. With this additional contribution, the induced electric polarisation in the nucleon is shown to extend to larger transverse distances.

Figure \ref{fig-world-betam} shows the values of the magnetic GP. Since this polarisability is smaller than the electric one, it is measured with a larger relative  error, and is sensitive to many systematic effects, such as the overall normalisation of the experiments, etc. One can say that the data globally reflect the expected $Q^2$-behaviour of $\bem$, similarly to the $\plt$ structure function in figure~\ref{fig-sf}. The shape of $\bem$ in the low-$Q^2$ region results from the compensation of two large contributions of opposite sign (dia- and paramagnetic) and it would be desirable to gain insight into this phenomenon.

More globally, the low-$Q^2$ region is the one expected to exhibit meson cloud effects \cite{Friedrich:2003iz} and this is one strong motivation for acquiring new and accurate VCS measurements. These mesonic effects have been known to be important in the nucleon polarisabilities,  since the first ChPT calculation of $\ale$ and $\bem$  in RCS~\cite{Bernard:1993bg}. They are also suggested by the large value of the electric polarisability 
%
%
mean square radius of the proton,  of $ (2.2 \pm 0.3)$ fm$^2$ \cite{Bourgeois:2006js}.

\subsection{Ongoing experiments}

We have seen that the $Q^2$-behaviour of the electric and magnetic GPs of the proton is definitely non-trivial. The existing data raise questions and call for new measurements. This is the aim of an ongoing VCS experiment at MAMI~\cite{Merkel:2009} which will measure the structure functions $\plltt$ and $\plt$ and the scalar GPs at three values of $Q^2$ yet unexplored: 0.1, 0.2 and 0.5 GeV$^2$. By combining in-plane and out-of-plane kinematics, this experiment is expected to provide new and accurate data and help to build a consistent picture of the scalar GPs of the proton.

\subsection{Other VCS observables}

Other VCS observables can be accessed, essentially by polarised experiments.
Using just a longitudinally polarised electron beam, one can measure a beam-spin asymmetry, as soon as the CM energy is above the pion threshold. This asymmetry is well understood in terms of the imaginary part of the VCS amplitude, and its physical content can be interpreted using the Dispersion Relation formalism.
Singly polarised $\epg$ experiments have been performed and interpreted along these lines \cite{Bensafa:2006wr,Sparveris:2008jx}, using the fact that the DR model works well in the region of the Delta resonance.


There are no measurements yet of the spin part of the structure functions
and of the spin GPs, despite their interest from the theory point of view.
Such experiments are especially difficult and challenging. A first VCS experiment with double polarisation has been undertaken at MAMI~\cite{Merkel:2000,Doria:sub}. Using a polarised electron beam and measuring the recoil proton polarisation, a double polarisation asymmetry is built. Below the pion threshold and using a LET~\cite{Vanderhaeghen:1997bx} one can in principle disentangle the various spin GPs from such data. However this pioneering experiment showed less sensitivity than expected, and only one new structure function could be extracted (see preliminary analyses in \cite{Janssens:2007th,Doria:2007th}).


Another interesting possibility would be to access the (pure spin) structure function $\ptt$, by measuring the  $\plltt$ structure function in the usual way (in an unpolarised experiment) and making a separation of $\pll$ and $\ptt$. This requires measurements at high and low $\epsilon$, i.e. a Rosenbluth-type technique. However the GP effect in the cross section scales with $\epsilon$, therefore at low $\epsilon$  it becomes hardly detectable, and no experiment of this type has been designed yet.


\section{Conclusion} \label{sec-concl}

Compton Scattering on the nucleon is an active field of research, which  has seen many new developments in the last decade. In RCS the electric and magnetic polarisabilities of the proton have been measured with a impressive accuracy, and in VCS a global picture of their $Q^2$-behaviour has begun to emerge. New experiments are planned or ongoing to pin down the unanswered questions: what are the values of the spin polarisabilities of the proton? what happens to the electric and magnetic GPs of the proton  at low $Q^2$? how do diamagnetism and paramagnetism compete versus $Q^2$? If these topics will receive new input in the near future, more difficult topics like the spin GPs of the proton, or neutron polarisabilities, will be addressed in a longer-term future.

The theoretical side is actively progressing too, including fully covariant ChPT, effective field theories, and light-front formalism. Altogether with new experimental data on Compton scattering, these developments will provide an important step forward in our understanding of the electromagnetic structure of the nucleon.
The infrastructure provided by both the theory group at KPH-Mainz and the A1 and A2 Collaborations at MAMI makes this laboratory a unique and exciting place in the world to lead the investigation of this physics.

\ \\


{\bf Acknowledgments}

This article is a follow-up of a presentation made at the ``CRC 443 Conclusive  Symposium'' in February 2011 in Mainz. We thank the Symposium organizers for giving us the opportunity to come up with this paper, the MAMI A1 and A2 Collaborations which are at the origin of many of the results presented here, and the KPH Hadron Physics Theory group for its support. E.J.D. gratefully acknowledges the support of a Postdoctoral Fellowship from the Carl Zeiss Stiftung.







\end{document}